\documentclass{article}

\usepackage[preprint]{neurips_2026}
\usepackage{amsthm}

\usepackage[utf8]{inputenc} 
\usepackage[T1]{fontenc}    
\usepackage{hyperref}
\usepackage{url}
\usepackage{booktabs}
\usepackage{amsfonts}
\usepackage{amsmath, amssymb}
\usepackage{nicefrac}
\usepackage{microtype}
\usepackage{pifont}
\newcommand{\cmark}{\ding{51}} 
\newcommand{\xmark}{\ding{55}} 
\usepackage{xcolor}
\usepackage{graphicx}
\usepackage{listings}
\usepackage{multirow}
\usepackage{float}
\usepackage{cite}
\usepackage{amsthm}

\title{Computer Science Conferences Should Require Nonrepudiable Experimental Results}

\author{
  Mamadou K. KEITA\textsuperscript{1},
  Christopher Homan\textsuperscript{1}
\\
\\
\textsuperscript{1}Rochester Institute of Technology
}

\begin{document}

\maketitle

\begin{abstract}

This position paper argues that computer science conferences should require tamper-evident, nonrepudiable attestations of experimental results. We name the underlying problem \textit{experiment nonrepudiation}: a compliant protocol must bind the numbers in a paper to an actual executed computation in a way the author cannot later alter or deny. The current system relies on self-reported checklists, optional code sharing, and author-controlled logging. None of these mechanisms answer the question a reviewer cannot check: did the code the paper describes produce the numbers the paper reports? We define the problem formally, state the security properties any compliant protocol must satisfy, and describe a threat model that includes attacks current approaches do not prevent. We frame the mechanism that provides this property as a \textit{proof-of-compute layer}, a general requirement that computed results be admitted as evidence only when accompanied by a verifiable record of the computation that produced them. To show that the problem is solvable, we built K-Veritas, the official reference implementation and public infrastructure for experiment nonrepudiation, written in Go, which produces signed reports without accessing training data. K-Veritas is in use today, with a public registry of sealed records; it is not final, and we intend to improve it through community collaboration. We call on conferences and the community to treat nonrepudiation as a major requirement and to help build an open, independent standard for it.
\end{abstract}

\section{Introduction}

Reproducibility is an important aspect of the scientific method. A result that cannot be independently verified contributes little to cumulative knowledge. In machine learning (ML), reproducibility has been a concern for over a decade, and the problem is not improving fast enough.

Kapoor and Narayanan~\citep{kapoor2022leakagereproducibilitycrisismlbased} surveyed the ML literature and found data leakage errors in 294 published papers across 17 scientific fields. Semmelrock et al.~\citep{semmelrock2024reproducibility} reviewed reproducibility barriers in ML research. They concluded that many papers are not reproducible in principle because of missing code, undocumented training conditions, or sensitivity to initialization. Fanelli's~\citep{Fanelli2009HowMS} meta-analysis of 18 survey studies found that 1.97\% of scientists admitted to fabricating, falsifying, or modifying data at least once, that up to 33.7\% admitted to other questionable research practices, and that 14.12\% reported seeing a colleague falsify data. Computer science is not exempt. Shepperd and Yousefi~\citep{Shepperd_2023} found that 2,816 of the 33,955 entries in the Retraction Watch database, roughly 8\%, are classified as computer science, that 56\% of retracted CS papers give little or no reason for the retraction compared to 26\% in other disciplines, and that a sample of 120 retracted CS papers still accumulated 672 post-retraction citations.

These are not edge cases. They are systematic failures. The publish-or-perish culture rewards novelty and strong numbers. Reviewers operate under strict deadlines with no budget to rerun experiments. As a result, reviewers take a paper's numbers on faith.

Conferences have responded with documentation-based measures. NeurIPS introduced a reproducibility checklist in 2021~\citep{neurips_checklist}. ICML adopted similar guidelines~\citep{icml2024guidelines}. ACM established artifact evaluation badges~\citep{acm_badges}. The ML Reproducibility Challenge invites volunteers to replicate accepted papers~\citep{pineau2021improving}. Tools like Weights \& Biases, MLflow, and Neptune log training runs. Moreover, pre-registration workshops~\citep{pmlr-v148-bertinetto21a} have piloted an alternative model in which experimental plans are reviewed before collecting results.

However, all of these measures share a common weakness: \textit{they are voluntary, self-reported, or post-hoc}. The checklist asks authors whether they disclosed training details, but it does not verify the answer. Artifact evaluation checks whether code runs, not whether it produced the reported numbers. Logging tools are author-controlled, so the author can modify or selectively share logs. Pre-registration commits to a plan before the experiment, yet it does not bind the reported numbers to the actual run. None of these mechanisms answer the simplest question: \textit{did the training run described in this paper actually produce the results this paper reports?}

Furthermore, the trust problem is not limited to results. ICML 2025 explicitly prohibits reviewers from using generative AI tools to write reviews or from entering any content from a submission into such a tool.\footnote{ICML 2025 Reviewer Instructions: \url{https://icml.cc/Conferences/2025/ReviewerInstructions}} The rationale is clear: the community cannot verify whether a review reflects real human assessment. The same logic applies, with equal or greater weight, to the experimental results that reviews are meant to evaluate. If fabricated reviews are a recognized threat worth prohibiting, fabricated results are a recognized threat worth verifying.

We argue that the underlying problem deserves a name and a definition. We call it \textit{experiment nonrepudiation}, borrowing the term from the security literature, where nonrepudiation means a party cannot later deny having acted~\citep{zhou1996fairnonrepudiation}. In empirical computer science, an author should not be able to alter, deny, or misrepresent what their computation actually produced, and the computation record should be independently verifiable.

\textbf{We argue that computer science conferences should require authors to submit tamper-evident, nonrepudiable attestations of their experimental results, generated by an independent, author-inaccessible protocol, that bind the reported numbers to the actual executed computations.}

The paper is organized as follows. Section~\ref{sec:problem} presents evidence that the reproducibility problem is structural and that current solutions are insufficient. Section~\ref{sec:trust} shows through two brief exercises that reported results alone cannot be trusted. Section~\ref{sec:nonrepudiation} defines experiment nonrepudiation as a problem class and states the security properties any compliant protocol must satisfy. Section~\ref{sec:threats} describes the threat model, including attacks current designs do not defeat. Section~\ref{sec:testbed} describes K-Veritas, the official implementation in use today. Section~\ref{sec:adoption} outlines a path to adoption. Section~\ref{sec:alternative} addresses alternative views. Section~\ref{sec:conclusion} concludes.

\section{The Verification Gap}
\label{sec:problem}

This section argues that a structural gap exists between what conferences ask for and what they actually verify. The problem is not new. Stodden et al.~\citep{stodden2016enhancing} identified reproducibility of computational methods as a systemic challenge across science, and Gundersen et al.~\citep{gundersen2022sources} cataloged specific sources of irreproducibility in machine learning, ranging from undisclosed random seeds to hardware sensitivity. We examine five categories of existing measures and explain why each still falls short.

\subsection*{Self-Reported Checklists}

NeurIPS requires a paper checklist that asks authors to confirm they disclosed training details, error bars, and compute resources~\citep{neurips_checklist}. The checklist is a step forward. It reminds authors to think about reproducibility. However, it is self-reported. An author who fabricated results can check ``yes'' on every item. Thus, the checklist verifies intention, not execution.

Gundersen and Kjensmo~\citep{gundersen2018state} surveyed 400 papers from AAAI and IJCAI and found that none documented all the variables required for reproducibility. Only 20--30\% of the necessary variables were documented per paper. The problem is not that researchers are careless. The problem is that checklists rely on voluntary disclosure, and voluntary disclosure is not enough.

Kapoor et al.~\citep{kapoor} developed REFORMS, a 32-item reporting checklist for ML-based science, built by consensus of 19 researchers across computer science, social science, and biomedicine. REFORMS is more comprehensive than any prior checklist. It covers data leakage, evaluation design, and uncertainty reporting. Nevertheless, it shares the same structural limitation: it asks authors to self-report. Someone who fabricated results can fill out the REFORMS checklist just as easily as the NeurIPS checklist. Better documentation standards help honest researchers avoid honest mistakes. They do not help when the mistake is deliberate.

Goldberg et al.~\citep{goldberg2024checklist} evaluated an LLM-based checklist assistant at NeurIPS 2024. The assistant helped authors verify checklist completion against the paper text. This is useful for catching honest omissions. However, it does not help when the omission is deliberate. The assistant checks whether the paper \textit{claims} to report error bars. It cannot check whether those error bars reflect real variance from real runs.

\subsection*{Artifact Evaluation}

ACM conferences offer artifact evaluation, where volunteers check that submitted code is documented, functional, and can produce results~\citep{acm_badges}. Papers that pass receive badges. This process has clear value. It incentivizes code sharing and catches broken pipelines.

However, artifact evaluation has three limitations. First, it is optional. Authors can decline without penalty. Second, for most venues, it occurs after acceptance, so it does not influence the accept/reject decision. Third, the Artifacts Evaluated Functional and Reusable badges certify that the code
is documented, complete, and runnable. They check whether the code \textit{can}
produce results, not whether it \textit{did} produce the particular numbers in the
paper. An author could submit working code that generates plausible outputs while
the paper reports numbers from a different, more favorable run.

Olszewski et al.~\citep{olszewski2023reproducibility} conducted a large-scale reproducibility study of ML papers at four top security conferences (USENIX Security, ACM CCS, IEEE S\&P, and NDSS) over a decade. They examined nearly 750 papers. Only 40\% included artifacts. Of the available artifacts, only 44\% ran successfully, meaning roughly 18\% of the studied papers produced working, available code. Most importantly, introducing Artifact Evaluation Committees at these venues produced no statistically significant improvement in artifact availability or functionality.

De Viti et al.~\citep{deviti2023hotosxixpanelreport} organized a panel at HotOS~2023 to discuss the future of artifact evaluation in systems research. The panel agreed that current AE goals are not aligned with community needs. Panelists agreed that AE should focus on ensuring artifacts are available and reusable for future work, not on verifying that exact numbers match. This is a reasonable position for artifact evaluation. However, it also means that artifact evaluation, even when functioning well, is not designed to verify results. It verifies usability.

\subsection*{Experiment Logging Platforms}

Tools like Weights \& Biases, MLflow, and Neptune log hyperparameters, metrics, and system information during training. These tools are valuable for internal experiment management. However, they are author-controlled. The author decides what to log, which runs to share, and whether to modify the logs before sharing. There is no independent verification. As a result, logs from these sources reflect \textit{what the author chose to show}, not \textit{what actually happened}.

\subsection*{Pre-Registration and Dataset Documentation}

Pre-registration has been proposed as an alternative publication model for ML, in which a paper is reviewed on the strength of its experimental plan before results are collected~\citep{forde2019scientificmethodsciencemachine, pmlr-v148-bertinetto21a, hofman2023preregistrationpredictivemodeling}. Pre-registration changes the review focus: reviewers assess the design, not the size of the numbers. It supplements any verification scheme. However, it is not a substitute. A pre-registered study still reports numbers after running, and those numbers are reported under the same system as any other paper. Pre-registration commits the \textit{plan}; it does not bind the \textit{execution}.

A parallel line of work has established standards for documenting ML pipeline inputs. Datasheets for Datasets~\citep{gebru2021datasheets} proposed that every dataset be accompanied by a structured document describing its motivation, composition, and collection process. These standards improve transparency about artifacts. However, they do not verify that the artifact described in the paper is the one actually used during the reported run.

\subsection*{Software Supply Chain Security}

Outside ML, the security community has built strong infrastructure to ensure the integrity of software artifacts. in-toto~\citep{torresarias2019intoto} cryptographically ensures the integrity of the software supply chain, recording signed attestations for each step of a build. Sigstore~\citep{newman2022sigstore} provides free, usable software signing for open-source releases. Both systems address a related but different problem: they bind a released artifact to a specified build process. However, neither binds a \textit{numeric result} (like an accuracy on a held-out set) to the \textit{computation} that produced it. This paper addresses that gap.

\subsection*{The AI Review Problem}

ICML 2025 prohibits reviewers from using generative AI tools to write reviews or from entering any submission content into such a tool. The reasoning is that the community has no way to verify whether a review reflects real human assessment. A review generated by a language model is indistinguishable (at some point) from a human-written one by inspection alone. Therefore, the community recognized this as a trust problem and responded with a prohibition.

The same problem applies to results. A result table generated by a language model asked to produce plausible benchmarks is indistinguishable from a table produced by an actual training run. The community's response to fabricated reviews is immediate and enforceable: submit one and face sanctions. By contrast, the community's response to fabricated results is a checklist.

\noindent
After all this analysis, the gap is simple. No existing mechanism at any major CS conference binds the numbers in a submitted paper to an actual executed computation in a tamper-evident, independently verifiable way. Checklists verify claims about the paper. Artifact evaluation verifies that code works. Logging platforms verify what authors share. Pre-registration commits to a plan. Software signing binds artifacts to builds. None of them bind reported results to real runs.

\section{Why Reported Results Alone Cannot Be Trusted}
\label{sec:trust}

Before we define the problem, two short exercises illustrate why verification matters.

Consider Table~\ref{tab:left} and Table~\ref{tab:right}. Both report results from fine-tuning a sentiment classification model on a standard benchmark. One table comes from a real training run. A language model generated the other, asked to produce plausible results for the same setup.

\begin{table}[h]
\centering
\begin{minipage}{0.47\textwidth}
\centering
\caption{Experiment A.}
\label{tab:left}
\scriptsize
\begin{tabular}{lcc}
\toprule
\textbf{Model} & \textbf{Acc.} & \textbf{F1} \\
\midrule
BERT-base & 91.8 & 91.5 \\
RoBERTa-base & 93.2 & 93.0 \\
DeBERTa-v3 & 94.1 & 93.8 \\
\midrule
Ours & \textbf{94.7} & \textbf{94.5} \\
\bottomrule
\end{tabular}
\end{minipage}
\hfill
\begin{minipage}{0.47\textwidth}
\centering
\caption{Experiment B.}
\label{tab:right}
\scriptsize
\begin{tabular}{lcc}
\toprule
\textbf{Model} & \textbf{Acc.} & \textbf{F1} \\
\midrule
BERT-base & 83.4 & 81.2 \\
RoBERTa-base & 76.9 & 74.6 \\
DeBERTa-v3 & 87.6 & 89.4 \\
\midrule
Ours & \textbf{91.3} & \textbf{87.0} \\
\bottomrule
\end{tabular}
\end{minipage}
\end{table}

Decide which table contains the real results. If you selected the left table, you are wrong. If you selected the right table, you are also wrong. Both tables were generated to make a point: \textit{you cannot distinguish real results from fabricated ones by looking at a table}. The numbers are plausible. The baselines are more or less consistent with published benchmarks. A reviewer reading either table in the context of a well-written paper would have no reason to suspect fabrication.

Setup descriptions are no harder to fabricate. A plausible-looking paragraph about the optimizer, learning rate schedule, batch size, and hardware can be produced without ever running a single batch. A sufficiently motivated reviewer could rerun the described configuration and compare, but no reviewer has the time or obligation to do this during a standard review cycle. The review process was not designed for it.

The review process evaluates the plausibility of results, not their authenticity. Therefore, the only reliable method is source verification: a tamper-evident record that binds reported numbers to actual computations, produced during execution by a process the author does not control.

\section{Experiment Nonrepudiation}
\label{sec:nonrepudiation}

This section defines the problem class this paper argues for.

\subsection*{Definition}

\textit{Experiment nonrepudiation} is the property that, for a given reported empirical result, a tamper-evident record binds the reported numbers to a specific executed computation, and the author cannot alter or deny this record after the fact.

The term is borrowed from security~\citep{zhou1996fairnonrepudiation}, where nonrepudiation classically means a party cannot later deny sending or receiving a message. Our use is similar, an author cannot later deny, nor can the author alter, the record of what their computation actually produced. Nonrepudiation is distinct from the adjacent concepts the community has already discussed. \textit{Reproducibility} asks whether someone else can rerun the experiment. \textit{Replicability} asks whether rerunning produces the same result. \textit{Provenance} asks where the data and code came from. By contrast, \textit{nonrepudiation} asks whether the reported result is tied to an actual execution the author cannot later misrepresent. Experiment nonrepudiation is the property, and the \textbf{\textit{proof-of-compute layer}} is what provides it. Just as physical science places an instrument between the experimenter and the claim, every empirical computational claim should be accompanied by a witness that the author does not control, so that a computed result is admitted as ``evidence'' only when it carries a nonrepudiable record of the computation that produced it.

\subsection*{Problem Specification}

We state the problem abstractly so that any compliant implementation can be evaluated against it.

\textbf{Inputs} A computation $C$ consisting of: executable code (source files, dependencies, framework versions), a configuration (hyperparameters, random seeds, data selections), a hardware environment (CPU, accelerators, memory), and a dataset $D$ (which is never exposed outside the author's machine). The computation produces a set of results from reported metrics $m_1, \ldots, m_k$ (accuracy, F1, loss, etc.). 

\textbf{Outputs} A signed attestation $A$ that ties together: a cryptographic digest of the code, a digest of the configuration, a fingerprint of the hardware environment, the reported metric values, a record of runtime telemetry (CPU time, memory, accelerator utilization), and a digest of the observed standard output. An independent party verifies the attestation against a public key.

\textbf{Required security properties} Any compliant protocol must satisfy the following.

\textit{Passivity} The observer must not modify the computation. Results must come from the author's run, not from an observer-modified version of it.

\textit{Data blindness} The observer must never access the dataset $D$. It may record size and pipeline structure, but not the data itself. Therefore, the protocol must not require authors to share sensitive or proprietary data.

\textit{Execution-binding} The reported metrics must be linked to the specific execution that produced them. Runtime telemetry must be linkable to real computation: a reported result on a large dataset trained on a GPU should show hardware activity consistent with that claim. A metric that appears without measurable computation is a flag.

\textit{Tamper-evidence} The attestation must be signed such that any modification to any field is detectable. Modifying a metric value, a hyperparameter, a timestamp, or even a single character of the recorded stdout must invalidate the signature.

\textit{Author-key separation} The signing key must not be held by the author. Without this property, the author can create arbitrary attestations. The key stays on an independent attestation service operated by a party with no stake in the paper's acceptance.

\textit{Independent verifiability} A separate tool, run by anyone (the conference, a reviewer, a future reader), must be able to validate the attestation without trusting the author. Verification is a public function of the signed record and a public key.

These properties are stated as protocol requirements, and any system that meets them is compliant.

\subsection*{Formalization}
To provide mathematical grounding for the properties above, we formalize the attestation procedure in a mathematical model. An attestation is a pair $A = (R, \sigma)$, where $R$ is the observer's
record of the execution (digests of the code, configuration, output streams,
and telemetry, together with the parsed metrics and the session time window)
and $\sigma = \mathrm{Sign}(\mathrm{Hash}(R))$
is the attestation service's signature over its digest, with the
public verification function $\mathrm{Verify}$. Experiment
nonrepudiation is then the requirement that, for every efficient adversary in
the covered class producing a candidate attestation $(R^*, \sigma^*)$,
 
\begin{equation}
\label{eq:enr}
\Pr\Big[\,\mathrm{Verify}(R^*, \sigma^*) = 1
\;\wedge\; R^* \neq R\,\Big]
\;\le\;
\varepsilon_{\mathrm{sig}} + \varepsilon_{\mathrm{hash}},
\end{equation}
where $R$ is the record of the execution the observer witnessed in
the sealed session, and $\varepsilon_{\mathrm{sig}}$ and
$\varepsilon_{\mathrm{hash}}$ are the probabilities of forging a signature or
finding a hash collision, both practically zero for the schemes used (RSA-PSS and SHA-256, in our case). In simple words, the only way to make a false record
pass verification is to break the signature scheme or the hash function.

\noindent
Although our examples are drawn from ML, experiment nonrepudiation is not specific to ML. The property applies to any empirical computational claim:  systems benchmarks, optimization results, computer-simulation-based scientific experiments, agent evaluations, etc. The protocol properties that make a compliant attestation work for ML work equally well for any field that produces empirical claims through computer-based pipelines. We view the scope as the largest class of problems at any conference or journal for which nonrepudiation is meaningful.

\section{Threat Models}
\label{sec:threats}

Tamper-evidence requires a meaningful threat model. We list the attacks a nonrepudiation protocol should consider, and for each one explain how the protocol responds.

\textbf{Text-level fabrication} The author edits numbers in the paper after the run, or invents numbers without running at all. The paper's claims are compared to the signed record at submission, and mismatches are detected.

\textbf{Log manipulation} The author modifies training logs after the run. A signed record with stdout digests freezes the logs at the time the session is sealed, so later edits invalidate the signature.

\textbf{Selective reporting} The author runs many times and reports only the favorable run. A signed session binds one run at a time, so a naive attacker submits an attestation of the chosen run and hides the others. Run anchoring closes most of this gap: every invocation under the observer, including failed and interrupted ones, is anchored at the attestation service when it ends, so the sealed record states how many runs occurred and how many were kept. What remains uncovered are runs performed entirely outside the observer. Pre-registration reduces this further, but nonrepudiation alone does not eliminate it.

\textbf{Fake training loops} The author writes a script that produces plausible metrics and telemetry without doing real work. A hardware-accountability layer flags superficial fakes: a paper claiming GPU training on a large dataset should show matching GPU activity and memory usage. An attacker running a compute-heavy script that produces chosen numbers does most of the work of real research.

\textbf{Operating system tampering} A compromised OS feeds false telemetry to a user-space observer. A modified kernel can return forged counters, or interpose library calls so the observer reads what the attacker wants. As a result, a user-space observer cannot prevent this.

\textbf{Firmware} A virtualized environment that lies about its hardware, or malicious firmware that misreports counters, is stronger still. A user-space observer cannot prevent this either.

\textbf{Attestation service compromise} If an attacker steals the signing key, they can produce valid attestations for anything. This is a governance and operations problem, not only a cryptographic one, and approaches such as federation, key rotation, and independent auditing handle it.

A software-only protocol handles text-level fabrication, log edits, naive selective reporting, and superficial fake training. However, it does not handle a privileged adversary with kernel or hardware access. Even so, the cost of fabrication changes. Without nonrepudiation, an author needs a text editor. With nonrepudiation, an author needs to run real computation or compromise a kernel.

\section{K-Veritas: The official tool and infrastructure}
\label{sec:testbed}

To show that the properties of Section~\ref{sec:nonrepudiation} are achievable in existing pipelines, we built K-Veritas, the official reference implementation of a proof-of-compute layer, written in Go, along with the public infrastructure around it: an attestation service, a web verifier, and a public registry of sealed records. Any other implementation meeting the required properties is equally valid, and we expect better designs to follow in the future.

The observer is a standalone compiled binary with no runtime dependencies. The author does not modify their code. Instead, they prefix their existing commands with \texttt{kveritas run}. The full workflow is two commands:

\lstset{language=bash}
\begin{lstlisting}
kveritas run --seal -- python train.py   # any command; writes report.pdf
kveritas verify report.pdf               # anyone, no account
\end{lstlisting}

The \texttt{kveritas} binary wraps the process at the OS level. It captures \texttt{stdout} and \texttt{stderr} non-blockingly (the author still sees their output), parses metrics from what the script prints, and hashes source files before and after each run. A background sampler records CPU time, memory usage, GPU usage, GPU memory, and disk I/O every second. At session close, \texttt{kveritas} computes a single canonical SHA-256 digest over the complete session (file hashes, stdout byte streams, parsed metrics, hardware samples, environment digest) and sends only that 64-character digest to the remote attestation service. As a result, the service never sees raw metrics, trajectories, or training data. It returns an RSA-PSS signature over the digest. The author never possesses the private key. Finally, the system produces a signed PDF report and a signed zip archive containing the source files that were present at execution.

Table~\ref{tab:report_fields} shows a snapshot of the fields captured in one published record, an independent reproduction of \texttt{karpathy/llama2.c} that trained a 15M-parameter Llama~2 on TinyStories\footnote{Record \texttt{kv:2610.00028v1}. The web-based verifier is available at \url{https://kveritas.org/verify} and the documentation at \url{https://kveritas.org/docs}. A K-Veritas technical report will accompany every version that brings major infrastructure updates.}. The full report, the sealed code bundle, and the signature are public, so every field below can be checked by anyone.

\begin{table}[h]
\centering
\caption{Snapshot of fields captured by K-Veritas.}
\label{tab:report_fields}
\scriptsize
\begin{tabular}{ll}
\toprule
\textbf{Field} & \textbf{Value} \\
\midrule
Session & \texttt{76a53d82}, sealed 2026-10-08 \\
Operating system & linux / amd64 \\
CPU & Intel Xeon W-2145 @ 3.70GHz, 16 cores \\
Memory & 31.0 GB \\
GPU & NVIDIA GeForce RTX 5060 Ti, 16311 MiB \\
Command & \texttt{bash kv\_run.sh} \\
Duration & 2 hours 23 minutes (06:18--08:41 UTC) \\
Invocations & 1 total: 1 kept, 0 failed, 0 interrupted; anchored at run end \\
Metrics & \texttt{train\_loss} 1.1897 and \texttt{val\_loss} 1.1907 at step 14000 (8 readings each); \\
 & \texttt{c\_inference\_tokens\_per\_second} 69.4935 \\
Seed committed & \texttt{seed:1337}, output line 1, 2026-10-08T06:18:35Z \\
Coherence (HMC) & Coherent (0.50): the telemetry channels move together, as one process \\
Telemetry & 7200 samples; CPU time 8580.0 s; peak memory 8.46 GB; peak GPU 99\% \\
Provenance & 66 tracked files, 4 snapshots, 0 changed at run end; 128 files read, 24 subprocesses \\
Withheld data & 50 TinyStories shards excluded by \texttt{.kveritasignore}, hashes still bound \\
Environment hash & \texttt{sha256:dd2958...} \\
Source code hash & \texttt{sha256:8303b2...} \\
Stdout hash & \texttt{sha256:3f9dbd...} \\
Metrics hash & \texttt{sha256:e6b3be...} \\
Code bundle hash & \texttt{sha256:ea76c3...} \\
Data hash (signed) & \texttt{sha256:d74a38...} \\
Digital signature & RSA-PSS-SHA256, 4096-bit, key fingerprint \texttt{eeac8165...} \\
\bottomrule
\end{tabular}
\end{table}

The stdout hash ties metric values to what the script actually printed. The source code hash ties the code to the executed version. Both are part of the signed data. We define a hardware-metric Consistency (HMC) score, reported as coherence, that measures whether the run's per-process telemetry channels co-fluctuate as reflections of a single computation. We view HMC as one signal among many that future implementations will refine.

K-Veritas does not prevent the OS-level and hardware-level attacks mentioned in Section~\ref{sec:threats}. Nevertheless, it stops casual and moderate fabrication, and it is ready for use today. It is not perfect: closing the remaining gaps, through hardware-backed attestation and federation across independent attesters, is the goal of collaboration with conferences, journals, universities, labs, and public institutions. Appendix~\ref{app:today} lists the features available at the time of this submission and shows fields from further sealed records in the public registry.

Beyond signing experiments, K-Veritas also instantiates \textit{agent nonrepudiation} (Section~\ref{sec:agent_repudiation}). Invoked as \texttt{kveritas init -{}-harness}, it records an autonomous agent session, or several agents running together under one record, as a hash-chained log of designated actions signed by the same attestation service, which again sees only hashes. A recording hook at each agent's tool boundary commits every designated action before it takes effect, and the record attributes each action to the agent that performed it, distinguishing independent top-level agents from their nested sub-agents.

\section{Agent non-repudiation}
\label{sec:agent_repudiation}
The same requirement extends from experiments to autonomous agents. When an AI agent acts on a user's behalf (a coding assistant, a tool-using research agent, an agent under evaluation), the analogous question is whether the actions the agent, or its operator, later claims were taken are the actions that were actually taken. We call this property \textit{agent nonrepudiation}. We formalize it below and deliberately leave the mechanism open. A hash chain is one way to obtain it; a sandbox with a reference monitor, digital signatures, a transparency log, or trusted-execution attestation are others.

\subsection*{Definition}

Fix a \textit{designation} $D$: a predicate that selects which actions must be accounted for, such as tool calls, file writes, messages, and sub-agent spawns. A system provides agent nonrepudiation with respect to $D$ if, for every executed action $x_k$ with $D(x_k)=1$, the record $L$ contains an entry $e_k$ that binds together (i) the identity $\alpha_k$ of the agent that performed $x_k$, (ii) its content $(\iota_k,\omega_k)$, the inputs and outputs observed at the time of the action, and (iii) the position of $x_k$ in the session order, such that after the session no party, including the agent itself and its operator, can deny or alter any of the three without a verifier $V$ rejecting and localizing the altered entry. This produces three guarantees: \textit{existence} (no party can deny that $x_k$ occurred or that $\alpha_k$ performed it), \textit{content} (no party can claim inputs or outputs other than $(\iota_k,\omega_k)$), and \textit{order} (no party can claim $x_k$ occurred at a different position). The content guarantee is over what was logged at the time of the action, not over re-execution, since agent outputs on stochastic hardware are not bit-reproducible.

\subsection*{Completeness}

Binding what is logged is not sufficient; a designated action must not be able to take effect without being logged. Completeness holds \textit{by construction} when no execution path lets a designated effect occur before its entry is committed, so that releasing the effect implies the entry exists. A policy that merely instructs the agent to log itself does not qualify, since a bug or a compromised operator could act and skip the entry.

\subsection*{Non-sequential and multi-agent sessions}

The session order need not be a single sequence, and there need not be a single agent. Several independent top-level agents may run under one record, as when an orchestrator drives a set of agents, and each may in turn spawn its own sub-agents. Concurrent actions form a partial order over the set of actors. Each designated action names the actor that performed it: an independent top-level agent, or a sub-agent scoped under the top-level agent that owns it. Spawn relations then induce, within each top-level agent, a tree of sub-agents to any depth, so the actors as a whole form a \textit{forest}. Because every entry carries its top-level agent and, for a spawn, the identifier of the sub-agent it created, the entire forest is reconstructed from signed facts; reattributing an action to a different agent, moving a sub-agent under a different parent, or hiding an agent's action is therefore detectable and localized. Agent nonrepudiation thus holds for non-sequential and multi-agent sessions, including many top-level agents that each have their own sub-agents, not only for a single linear one.

We instantiate these properties in K-Veritas: a session begins with a server-signed genesis that binds $D$; each designated action becomes a hash-chained entry; and a signed seal closes the chain, while the recording hook ensures completeness by construction. Verification recomputes the chain and localizes the first failing entry. Appendix~\ref{app:agent_construction} gives the hash-chained construction.

\section{Path to Adoption}
\label{sec:adoption}

An independent non-profit organization with no institutional affiliation and no restrictive financial ties to any research lab, company, or university should maintain nonrepudiation of experimental results as an open standard. The model is similar to OpenReview~\citep{openreview}, which provides peer review infrastructure as a community resource without being owned by any single institution. The governance model is independent by design: no single entity should control the verification standard that the community relies on. K-Veritas is run on this principle: an open protocol, no ownership shares, and support from conferences, journals, universities, labs, governments, and people committed to good science.

For adoption, we propose three phases.

\paragraph{Phase 1: Voluntary} Conferences offer nonrepudiation attestations as an optional submission component. Papers that include verified reports receive a visible badge. Reviewers can check reports through a web verifier without installing software.

\paragraph{Phase 2: Expected} Conferences make attestations expected but not required, similar to how code submission evolved at NeurIPS. The review form notes absences. Verification is integrated into the submission portal so it happens automatically at upload time.

\paragraph{Phase 3: Required} Conferences require attestations for all empirical papers. Papers without them are desk-rejected or flagged for additional scrutiny. Attestation status becomes part of the standard reviewer information.

Phase 3 is the end goal. Reaching it requires a mature protocol, broad tool support, federation across multiple attestation providers, and community consensus. We estimate 3--8 years from initial adoption. Phase 1 can start now: the CLI, the web verifier, and the public registry of records already exist. We invite conferences to adopt Phase 1 and developers to contribute framework support and alternative verification backends.

\section{Alternative Views}
\label{sec:alternative}

We present six objections to our position and respond to each.

\paragraph{``This adds overhead to an already slow process.''}
We partially agree. Any new requirement adds friction. However, integrating a compliant observer is about wrapping an existing command with a prefix. The report is generated automatically. The overhead is comparable to adding a logging library. Furthermore, the cost of \textit{not} verifying results (wasted follow-up research, retracted papers, weakened trust) is far greater than the cost of wrapping a training loop.

\paragraph{``Motivated cheaters will find workarounds.''}
This is true, and we do not claim otherwise. Section~\ref{sec:threats} lists the attacks a software-only scheme does not defeat. The point is not perfection. The point is that fabrication costs change. Without nonrepudiation, fabrication requires only a text editor. With it, fabrication requires running real computation or compromising a kernel.

\paragraph{``Pre-registration already solves this.''}
Pre-registration and nonrepudiation are complementary, not substitutes. Pre-registration commits to an experimental plan before the experiment runs; it changes the incentive structure of peer review. By contrast, nonrepudiation links the reported numbers to an actual run; it changes the evidentiary status of reported results. A conference can require both. Neither alone closes the gap the other fills.

\paragraph{``Industry labs may not easily comply because of concerns.''}
A tiered metadata schema handles this. A minimal tier requires only final metrics, timestamps, framework versions, and random seeds. It does not require GPU model names, internal infrastructure details, or anything that reveals proprietary architecture choices. Labs that want stronger verification opt into higher tiers. Labs that cannot disclose hardware details comply at the minimal tier. Therefore, partial compliance is better than no compliance.

\paragraph{``This punishes honest researchers for the misconduct of a few.''}
Nonrepudiation \textit{protects} honest researchers. In a setting where all results are verified, honest work carries credibility. Right now, an honest researcher's results have the same evidentiary status as a dishonest researcher's results: \textbf{unverified}. Nonrepudiation changes this. As a result, verified results are more trustworthy, which benefits the researchers who produced them honestly.

\paragraph{``A centralized attestation server may have too much power regarding scientific legitimacy.''}
This objection is serious. A single server that decides whether results are valid becomes a single point of failure and control. Two design choices address the concern. First, governance: the organization must be explicitly independent, with a public protocol specification and multiple compliant implementations. Second, advisory status: the verification result informs judgment; it does not replace it. A paper without an attestation can still be accepted. A paper with one can still be rejected. Federation across multiple independent attestation providers is the long-term answer, and we call on the community to help design it.

\section{Conclusion}
\label{sec:conclusion}

We argued that computer science conferences should require nonrepudiable experimental results. We named the problem, defined it as a set of security properties that any compliant protocol must satisfy, and presented a threat model that is honest about what software-only schemes defeat and what they do not. More broadly, we proposed a proof-of-compute layer as the general mechanism for providing experiment nonrepudiation across computational science. We showed through two brief exercises that reported results alone cannot be trusted. We described K-Veritas as the official implementation and infrastructure, ready for use today and built to keep improving, and indicated hardware-backed attestation as the next step.

We believe that nonrepudiation should apply universally. Verification does not depend on who you are, where you work, or how famous your lab is. ICML already prohibits AI-generated reviews because the community cannot verify that a review reflects real human assessment.\footnote{ICML 2025 Reviewer Instructions: \url{https://icml.cc/Conferences/2025/ReviewerInstructions}} The same logic applies to results. If the community accepts that fabricated reviews are a threat worth addressing by policy, it should accept that fabricated results are a threat worth addressing by nonrepudiation.

Trust in science is built on evidence. A tamper-evident, independently verifiable attestation is stronger evidence than a table in a paper. The protocol properties that make nonrepudiation applicable to ML are not unique to ML, and we expect the framework to generalize to any field in which computational pipelines produce empirical results.

We invite the community, organizations, and conferences to help build it.

\bibliography{custom}
\bibliographystyle{plain}

\clearpage
\appendix

\section{Comparison with Established Approaches}
\label{app:comparison}

\begin{table}[h]
\centering
\caption{Comparison of nonrepudiation with existing reproducibility mechanisms.}
\label{tab:comparison}
\small
\begin{tabular}{lcccccc}
\toprule
\textbf{Property} & \textbf{Checklist} & \textbf{Artifact} & \textbf{W\&B/} & \textbf{MLRC} & \textbf{Pre-reg.} & \textbf{Nonrepud.} \\
 & & \textbf{Eval.} & \textbf{MLflow} & & & \textbf{(K-Ver.)} \\
\midrule
Binds result to run & \xmark & \xmark & \xmark & Partial & \xmark & \cmark \\
Tamper-evident & \xmark & \xmark & \xmark & \xmark & \xmark & \cmark \\
Pre-review & \cmark & \xmark & N/A & \xmark & \cmark & \cmark \\
No data access required & \cmark & \xmark & \xmark & \xmark & \cmark & \cmark \\
Author cannot modify record & \xmark & \xmark & \xmark & \cmark & Partial & \cmark \\
Automated verification & \xmark & \xmark & \xmark & \xmark & \xmark & \cmark \\
Universal (all submissions) & \cmark & \xmark & \xmark & \xmark & \xmark & \cmark$^*$ \\
\bottomrule
\multicolumn{7}{l}{\scriptsize $^*$ When adopted as required and currently proposed as voluntary first (Phase 1).}
\end{tabular}
\end{table}

\section{Limitations}
\label{app:limitations}

We acknowledge five limitations.

First, nonrepudiation binds reported numbers to an actual execution. However, it does not verify that the execution was well-designed. A poorly controlled experiment with real, verified numbers is still a poor experiment.

Second, the tamper-evidence guarantee depends on the attestation service's security and the security of its signing key. If an attacker compromises the service, they can fabricate attestations. This risk exists in any security protocol and requires careful infrastructure management, federation across multiple independent attesters, and regular key rotation.

Third, nonrepudiation requires authors to use a compliant implementation. If an author refuses, the protocol generates no attestation. As a result, the protocol works only when conferences make compliance mandatory or strongly encouraged.

Fourth, a software-only observer does not defeat OS-level or hardware-level adversaries (Section~\ref{sec:threats}). Hardware-backed attestation is the planned next step for K-Veritas and the way forward for high-assurance submissions.

Fifth, production deployment at conference scale requires institutional infrastructure: persistent session storage, rate limiting, key rotation, and auditing. These are operational requirements, not protocol limitations.


\section{Agent nonrepudiation hash-chained details}
\label{app:agent_construction}

We record a session as an append-only chain. A server-signed \textit{genesis} binds the designation $D$, the session identity, and a start time; its hash $\ell_0$ is the chain root. Each designated action becomes an entry $e_j$ containing its actor $\alpha_j$, the digests of its content, and a link $\ell_j = \mathrm{Hash}(e_j \parallel \ell_{j-1})$, where $\mathrm{Hash}$ is a collision-resistant hash. At session close, the server signs the final link $\ell_N$. A verifier $V$ recomputes the genesis hash, checks its signature, walks the chain recomputing each $\ell_j$, and checks the signature on $\ell_N$.

Under collision resistance of $\mathrm{Hash}$, this yields the three properties of Section~\ref{sec:agent_repudiation}. Fabricating an entry $e_j' \neq e_j$ with the same link is infeasible, so content and identity are bound; deleting, inserting, or relocating an entry changes some $\ell_j$ on which every later link and the signed $\ell_N$ depend, so order is bound, and $V$ rejects and reports the first inconsistent entry. Attribution is bound the same way: each entry names its top-level agent and its acting agent, and a spawn entry names the identifier of the sub-agent it created, so the whole forest of agents, several independent top-level agents each with their own sub-agents, is a function of the signed record and cannot be silently rewritten. Because the genesis binds $D$ before the session runs and the recording hook commits each entry before its action takes effect, the guarantee is exactly as broad as $D$ and admits no unlogged designated action.

\section{K-Veritas Today}
\label{app:today}

As of the date of this submission, K-Veritas provides the components listed in Table~\ref{tab:features}. The protocol is open and non-proprietary; verification requires no account and no contact with the author. Table~\ref{tab:real_records} shows fields taken from two further sealed records in the public registry~\footnote{\url{https://kveritas.org/records}}: one official implementation of a research paper and one independent reproduction of third-party code. At the time of writing, the registry holds 43 records (\texttt{kv:2610.00001} to \texttt{kv:2610.00043}) across computer vision, databases, machine learning, natural language processing, networking, programming languages and compilers, robotics, security and cryptography, and systems and performance. We plan to release a K-Veritas technical report for every version that includes major infrastructure updates.

\begin{table}[h]
\centering
\caption{K-Veritas components available at the time of submission.}
\label{tab:features}
\scriptsize
\begin{tabular}{lp{8.6cm}}
\toprule
\textbf{Component} & \textbf{What it does} \\
\midrule
CLI (\texttt{run}, \texttt{seal}, \texttt{verify}) & Wraps any command in any language; passive observation, no code changes; writes a signed PDF report. \\
Recorded shell & Records an interactive session as one sealed record. \\
Attestation server & Signs the session digest with an RSA-PSS 4096-bit key the author never holds; sees only hashes. \\
Origin ledger & Public ledger of who signed what, so a report can be checked against the server's record. \\
Run anchors & Every invocation is anchored at the server when it ends, including failed and interrupted runs; the report states how many runs occurred and how many were kept. \\
Coherence check (HMC) & Tests whether per-process telemetry channels move together as one computation. \\
Compute check & Compares reported results against the hardware activity observed during the run. \\
Code provenance and privacy & Hashes tracked files at run start and run end; files listed in \texttt{.kveritasignore} are withheld but their hashes are still bound; data never leaves the machine. \\
Single-file proofs & One report file carries everything a verifier needs. \\
Model and data artifacts & Binds model checkpoints and dataset hashes into the record without disclosing their contents. \\
Agent harness & Hash-chained, server-signed log of designated agent actions (Section~\ref{sec:agent_repudiation}). \\
Web verifier & \url{https://kveritas.org/verify}; checks a report in the browser, no account needed. \\
Public records & Submit, update, request changes, and cite sealed records with \texttt{kv:} identifiers at \url{https://kveritas.org/records}. \\
Documentation and tutorials & Install, first report, instrumentation, and tutorials for PyTorch, PyTorch Lightning, Keras, scikit-learn, XGBoost, LightGBM, and Hugging Face. \\
Platforms & Standalone binaries with no runtime dependencies; see the platform page in the documentation. \\
\bottomrule
\end{tabular}
\end{table}

\begin{table}[h]
\centering
\caption{Fields from two further sealed records in the public registry. Both reports are fully verifiable at \url{https://kveritas.org/verify}.}
\label{tab:real_records}
\scriptsize
\begin{tabular}{lll}
\toprule
\textbf{Field} & \textbf{S\ensuremath{\varepsilon}b\ensuremath{\varepsilon}n evaluation} & \textbf{Mamba-370M zero-shot} \\
\midrule
Record & \texttt{kv:2610.00041v1} (1 of 22 reports) & \texttt{kv:2610.00033v1} \\
Purpose & Official implementation, research paper & Independent reproduction \\
Code & Author's own & \texttt{state-spaces/mamba} \\
Sealed & 2026-10-12 & 2026-10-10 \\
GPU & NVIDIA L4, 23034 MiB & NVIDIA GeForce RTX 5060 Ti, 16311 MiB \\
CPU & Intel Xeon @ 2.20GHz (8 cores) & Intel Xeon W-2145 @ 3.70GHz (16 cores) \\
Duration & 1.8 seconds & 5 minutes 15 seconds \\
Reported results & \texttt{phi} 0.500696; \texttt{phi\_bam} 0.542398 & LAMBADA acc 0.556; PIQA acc 0.695 \\
Telemetry samples & 8 & 3136 \\
CPU time / peak memory / peak GPU & 0.0 s / 0.00 GB / 0\% & 374.2 s / 37.67 GB / 97\% \\
Coherence (HMC) & Not enough telemetry to judge & Coherent (0.36) \\
Tracked files / withheld & 27 / none & 84 / 4 CUDA sources (hashed) \\
Invocations (kept / failed) & 1 (1 / 0), anchored at run end & 1 (1 / 0), anchored at run end \\
Source code hash & \texttt{sha256:4c9693...} & \texttt{sha256:ab9c9a...} \\
Stdout hash & \texttt{sha256:207038...} & \texttt{sha256:419bf0...} \\
Data hash (signed) & \texttt{sha256:049664...} & \texttt{sha256:1ca0f6...} \\
Digital signature & RSA-PSS-SHA256, 4096-bit & RSA-PSS-SHA256, 4096-bit \\
\bottomrule
\end{tabular}
\end{table}

The left column is worth reading for what it does not claim. The run is an evaluation pass of a few seconds, and the report says so: the telemetry is too thin to judge coherence. The seal, the anchors, and the hashes still hold, so the numbers remain bound to the code that printed them, but the execution-binding signal is absent, and the report states this rather than hiding it. Reporting a record's limits is part of what makes it evidence.

\section{Reflection on Experiment Nonrepudiation}
\label{app:reflection}

It is not rare to see reviewers complain about AI-generated papers, or about misconduct in computational science more generally. Many even acknowledge that they do not receive code with a submission or cannot rerun the experiments. We do not believe a single infrastructure or method solves this problem. What it asks for is an epistemic change in what should count as ``evidence'' in computational science; introducing experiment nonrepudiation and the K-Veritas infrastructure are steps toward solving portions of it, and larger portions in the future.

We expect people to be reticent, or even to call this (K-VERITAS) ``slop'' without trying to understand it or to help improve it. But we believe that experiment nonrepudiation clearly defines the limits of what current reviews can treat as evidence in computational science, as Turing's halting problem defined the limits of what can be decided~\citep{turing_halting}, and that it is a good starting point for addressing them. We do not know how big this could be. It may give rise to a whole new field, or to stronger nonrepudiation approaches; it may never reach any impact at all. Nevertheless, we know this for sure: if computation cannot produce evidence that stands on its own, results become simple claims. Claims that cannot be checked open uncertainty. Uncertainty that costs nothing spreads. And slowly, through compounding effects, trust fades from research.


\newpage

\end{document}